\documentclass[conference, a4paper]{IEEEtran}
\usepackage{blindtext}
\usepackage{url}
\usepackage{amsmath}
\usepackage[shortcuts]{extdash}
\usepackage{algorithm}
\usepackage{algpseudocode}
\usepackage{tabularx}
\usepackage{float}
\usepackage[normalem]{ulem}

\usepackage{scalerel}
\usepackage{tikz}
\usepackage{comment}
\usetikzlibrary{svg.path}
\definecolor{orcidlogocol}{HTML}{A6CE39}
\tikzset{
  orcidlogo/.pic={
    \fill[orcidlogocol] svg{M256,128c0,70.7-57.3,128-128,128C57.3,256,0,198.7,0,128C0,57.3,57.3,0,128,0C198.7,0,256,57.3,256,128z};
    \fill[white] svg{M86.3,186.2H70.9V79.1h15.4v48.4V186.2z}
                 svg{M108.9,79.1h41.6c39.6,0,57,28.3,57,53.6c0,27.5-21.5,53.6-56.8,53.6h-41.8V79.1z M124.3,172.4h24.5c34.9,0,42.9-26.5,42.9-39.7c0-21.5-13.7-39.7-43.7-39.7h-23.7V172.4z}
                 svg{M88.7,56.8c0,5.5-4.5,10.1-10.1,10.1c-5.6,0-10.1-4.6-10.1-10.1c0-5.6,4.5-10.1,10.1-10.1C84.2,46.7,88.7,51.3,88.7,56.8z};
  }
}

\newcommand\orcidicon[1]{\href{https://orcid.org/#1}{\mbox{\scalerel*{
\begin{tikzpicture}[yscale=-1,transform shape]
\pic{orcidlogo};
\end{tikzpicture}
}{|}}}}
\usepackage[hidelinks]{hyperref}

\begin{document}

\title{Performance Optimization of High-Conflict Transactions within the Hyperledger Fabric Blockchain}

\author{
Alexandros Stoltidis \orcidicon{0009-0001-3959-5916} and Kostas Choumas \orcidicon{0000-0001-5237-7254} and Thanasis Korakis \orcidicon{0000-0003-0162-335X}\\
\normalsize
Dept. of ECE, University of Thessaly, Volos, Greece\\
Email: {stalexandros, kohoumas, korakis}@uth.gr
}

\maketitle

\begin{abstract}

\textit{Hyperledger Fabric (HLF)} is a secure and robust \textit{blockchain (BC)} platform that supports high-throughput and low-latency transactions.
However, it encounters challenges in managing conflicting transactions that negatively affect throughput and latency.
This paper proposes a novel solution to address these challenges and improve performance, especially in applications incorporating extensive volumes of highly conflicting transactions. 
Our solution involves reallocating the \textit{Multi-Version Concurrency Control (MVCC)} of the validation phase to a preceding stage in the transaction flow to enable early detection of conflicting transactions.
Specifically, we propose and evaluate two innovative modifications, called \textit{Orderer Early MVCC (OEMVCC)} and \textit{OEMVCC with Execution Avoidance (OEMVCC-EA)}.
Our experimental evaluation results demonstrate significant throughput and latency improvements, providing a practical solution for high-conflict applications that demand high performance and scalability.
\end {abstract}


\section{Introduction}
BC is a \textit{Distributed Ledger Technology (DLT)} \cite{Sunyaev2020}, maintaining an immutable, distributed, and decentralized ledger of transactions across a network of nodes that reach a consensus on their states without a central authority \cite{cachin2017blockchain}.
DLT encompasses several distributed design paradigms and their underlying technologies, determining the design of the BC network.
The two most prominent paradigms are permissionless (public) and permissioned (private), which govern the access control of participants on the BC \cite{bakos_halaburda_2021}.
Contrary to permissionless BC networks like Bitcoin \cite{bitcoin} and Ethereum \cite{ethereum}, our research focuses on permissioned BC systems. 
These are less computationally intensive and not publicly accessible, necessitating cryptographic authentication and authorization from participants.
Examples of software platforms implementing such BC systems include HLF \cite{hlf}, Quorum \cite{quorum}, and Corda \cite{corda}.

Our research focuses on HLF, an integral component of the Hyperledger project, an open-source collaborative initiative facilitated by the Linux Foundation.
The HLF network contains various types of distinct nodes, including \textit{peers} and \textit{orderers}, and follows an \textit{execute-order-validate (EOV)} transaction flow model \cite{hlf-overview-model}.
Peer nodes execute and endorse individual transactions and validate and commit transactions stored within blocks to the ledger, encompassing the \textit{world state}, the \textit{log history}, and the \textit{BC}.
Orderer nodes incorporate consensus mechanisms, orchestrating transaction ordering and block creation.

We focus on maximizing transaction throughput while minimizing latency on applications where high-conflict transactions compete to modify the same assets within HLF.
Transaction throughput is the rate at which transactions are committed to the ledger, and latency is the time it takes since a client submits a transaction proposal till the transaction gets committed \cite{throughput-latency}.
We thoroughly examine and evaluate the different stages a transaction undergoes in the HLF transaction flow, considering the EOV transaction flow. 
Based on our comprehensive analysis, we propose and assess various optimizations in the execution and validation phases.

The remainder of this paper is structured as follows: Section \ref{sec:system} provides a more compact overview of the architecture and transaction flow of HLF. It delves deeper into its working intricacies and investigates the different stages transactions undergo, highlighting their suboptimal performance under specific circumstances.
Section \ref{sec:related} explores the existing literature regarding transaction throughput and latency optimizations.
In Section \ref{sec:solution}, we introduce our proposed solution, while in Section \ref{sec:evaluation}, we describe our evaluation framework and the conducted experiments alongside their evaluation results.
Finally, Section \ref{sec:conclusion} summarizes the content of this paper and discusses the future research directions. 

\section{System Description - The HLF Architecture} \label{sec:system}

In our research, we employ the latest HLF architecture, containing clients, peers, the \textit{gateway (GW)} \cite{gateway}, the orderers, the \textit{certificate authority (CA)}, and the \textit{membership service provider (MSP)}.
The CA and the MSP are responsible for the \textit{Public Key Infrastructure (PKI)}, detecting the permissions over resources and access to information that actors have in a BC network.
The GW is integrated within each peer, providing a layer of abstraction to the clients to eliminate concerns regarding the underlying network infrastructure.

Within an HLF network, a peer can function as either an \textit{endorsing peer} or a \textit{non-endorsing peer}.
Endorsing peers execute and endorse transactions using \textit{smart contracts} or \textit{chaincodes}, while non-endorsing peers only validate and commit blocks of transactions to the ledger.
Smart contracts establish the executable logic, while chaincodes bundle and instantiate multiple smart contracts on endorsing peers, making these terms often interchangeable.

The ordering service consists of orderers, incorporating a consensus mechanism (Apache Kafka/Zookeeper \cite{kafka} or Raft \cite{raft}) to establish a total order for the received transactions. 
The orderers create blocks from the ordered transactions and broadcast them to all the peers in the BC network using a gossip protocol. 
Ultimately, communication within the HLF network occurs through \textit{channels} \cite{8525394}. 
Each channel is a private communication link between two or more network members, facilitating private and confidential transactions.

\subsection{The EOV Transaction Flow}
\begin{figure}[htbp]
\vspace{-0.4cm}
\centerline{\includegraphics[width=0.5\textwidth]{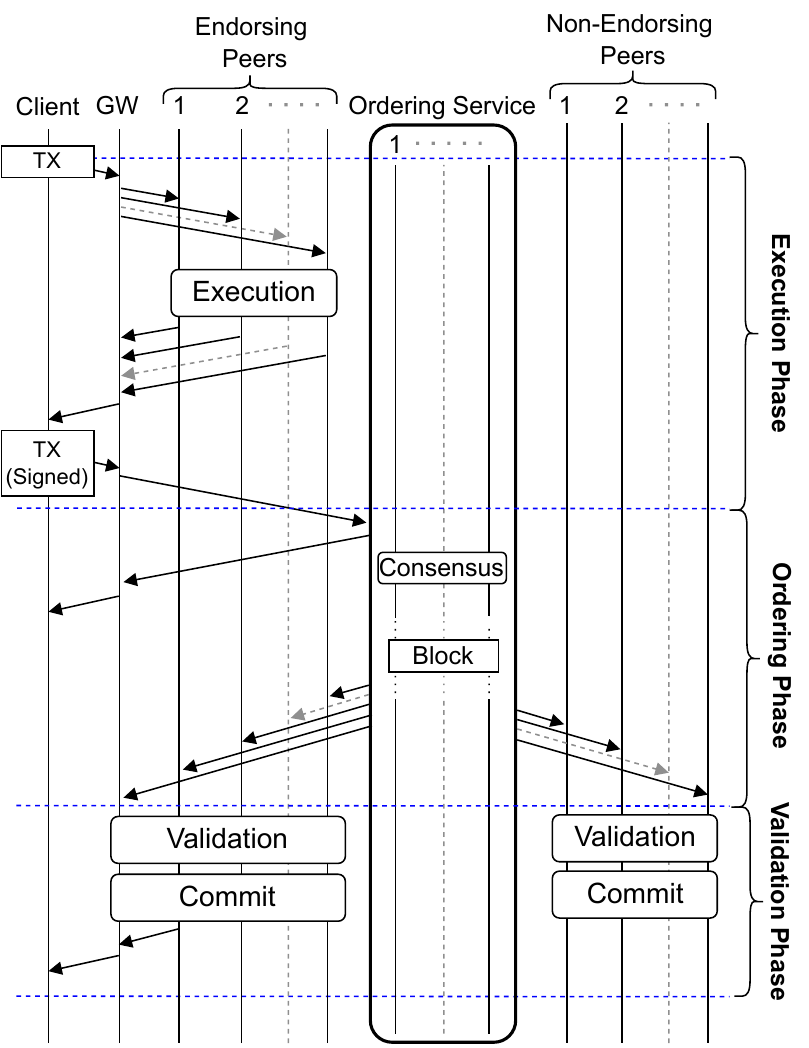}}
\vspace{-0.2cm}
\caption{The EOV Transaction Flow.}
\label{fig:eov}
\end{figure}

Figure~\ref{fig:eov} depicts the EOV transaction flow, outlining the comprehensive sequence of operations that every transaction undergoes. 
Specifically, it details the three distinct phases within the HLF network, including execution, ordering, and validation.

\subsubsection{Execution}
In the execution phase, a client creates a transaction proposal and sends it to the GW, residing on an endorsing or a non-endorsing peer.
In the case of Figure~\ref{fig:eov}, the GW is on a non-endorsing peer and forwards the received proposal to the endorsing peers, specified by the endorsement policy established during channel initialization.
Each endorsing peer executes the transaction in an isolated simulation environment against a copy of its world state, a key-value database that stores the information affected by the transactions and computes the read-write set of the transaction. 
The read set comprises the keys accessed and their version numbers, while the write set contains the keys that the transaction will modify alongside their new values. 
If no errors occur during the transaction execution, the peer sends back an endorsement to the GW.
After gathering sufficient endorsements, the GW sends the endorsed transaction proposals in a transaction envelope to the client for signing.
The client submits the signed transaction envelope to the GW, which then forwards it to the ordering service.

\subsubsection{Ordering} 
Upon receipt of the signed transaction envelope, the ordering phase commences.
Firstly, the ordering service notifies the client via the GW that it will eventually include the received transaction envelope in a block. 
The orderers within the ordering service utilize a consensus mechanism alongside a couple of predefined parameters to order the received transactions into blocks.
These parameters encompass the maximum transaction limit within a block, known as block size, and the block generation interval, utilized to produce blocks when they contain fewer transactions than the block size, referred to as the block interval.
Eventually,  the ordering service broadcasts the generated blocks to all peers for validation and subsequent commitment.

\subsubsection{Validation} 
The peers receive blocks either from the ordering service or from other peers through the gossip protocol.
In the validation phase, each peer validates the transactions in each received block, commits the blocks to the BC, and updates the world state with the valid transactions. 
Finally, one of the peers notifies the client via the GW that it has successfully committed the transaction to the ledger.


\subsection{Processes currently included in Validation}
Once a peer receives a block, it validates it before committing it to the ledger \cite{mvcc-vscc}.
The peer first verifies the syntactic signature of the received block and sends it through a pipeline of distinct operations to validate each transaction individually.  
After syntactically validating each transaction within the block, it executes the \textit{Validation System ChainCode (VSCC)} to ensure that endorsements adhere to the specified chaincode endorsement policy.
A transaction might fail VSCC if it lacks adequate endorsements, encounters issues with endorser signatures, or experiences mismatches in the read-write set versions collected by different peers with different ledger states.
The latest versions of HLF also incorporate VSCC on the GW, invalidating transaction envelopes before dispatching them to the ordering service and notifying the client earlier within the transaction flow.
Transactions failing VSCC are designated as invalid to prevent their commitment to the ledger.

Transactions that successfully pass VSCC go through MVCC, mitigating the double-spending problem.
MVCC ensures that valid transactions have no read-write conflicts by comparing the versions of keys in the read set at the execution phase with the versions of these keys on the world state database. 
MVCC invalidates transactions upon detecting a read conflict or a phantom read.
The former arises as every operation at the execution phase reads the current version of a key alongside its value from the world state, while every successful write operation increments the version number of the key upon commitment.
As a transaction progresses from the execution phase, concurrent transactions may get committed, incrementing the version numbers and altering the world state.
Consequently, when a transaction reaches the validation stage, its version numbers and perspective of the world state may have changed, which results in an MVCC read conflict.
Phantom read conflicts occur during range queries by checking the entire range of keys to ensure no insertions, deletions, or updates have occurred within it.
If any alteration within the range is detected, MVCC flags the transaction as invalid, prompting its termination.
Finally, the peer commits the block to the ledger, incorporating the valid transactions into the \textit{world state} and \textit{log history} and the block into the \textit{BC}. 

The problems emerge when multiple clients with conflicting interests introduce transactions referencing identical keys.
Peers do not detect potential conflicts in the early transaction stages; instead, a transaction is marked invalid during the validation phase, resulting in inefficient performance regarding transaction throughput and latency.
Shifting a part of the validation phase, particularly the MVCC process, to earlier stages of the transaction flow would enhance resource efficiency, thereby boosting the overall system performance.

\section{Related Work} \label{sec:related}
As a recently developed system, HLF undergoes significant architectural innovations.
Current research efforts aim to enhance system performance by maximizing transaction throughput and resource utilization while minimizing latency.
However, limited work on early detection mechanisms exists for conflicting transactions in HLF.

The existing literature primarily addresses mitigating early detection during the execution or ordering phases.
The study presented in \cite{8939202} strives to enhance read operations by distinguishing them from write transactions during the execution phase.
As read transactions operate differently from write transactions, clients can receive their results during the endorsement, bypassing the rest of the transaction flow. 
This process minimizes latency and block sizes, ultimately maximizing the overall throughput.
In \cite{xu2019locking}, a locking-based mechanism detects conflicting transactions, mitigating the effects of such transactions in high-concurrency applications. 
However, this mechanism relies on synchronized access to a trusted distributed locking service, necessitating extensive coordination and utilization of network resources, creating significant overheads, and making the mechanism unsuitable for delay-critical applications.
Research in \cite{emvcc1} and \cite{emvcc2} investigates the concept of performing part of MVCC on peers following the endorsement execution.
Each peer incorporates a local cache, storing the read-write sets of the endorsed transactions. 
The authors propose to abort transactions when they depend on a key stored in the cache since they will inevitably fail the MVCC in the validation phase.
In deployments featuring multiple peers are available, valid transactions detected at the execution phase may be invalid in the validation phase, according to their position within the block, requiring validating these transactions twice.
Consequently, transactions initially considered invalid during the execution phase may become valid, as preceding valid transactions could be invalid during the validation phase.

Introducing part of the validation during the execution phase of the transaction flow may yield inconsistent and invalid results.
Contrary to transactions within blocks, the arrival of transaction proposals on peers is not totally ordered, potentially leading to transaction invalidation that could otherwise be deemed valid.
Several research papers have proposed mechanisms to detect conflicting transactions in the ordering phase instead of the execution phase. 
In \cite{Sharma2019BlurringTL} and \cite{ruan2020transactional}, the authors have proposed a mechanism to reduce the MVCC failure rate by using a conflict graph to reorder transactions at the ordering stage and abort conflicting transactions.
The research paper in \cite{ordering-mvcc} investigates a pipeline to reduce conflicting transactions in the ordering service. 
The orderers group the transactions into blocks to facilitate parallel processing, filter out obsolete transactions, and prioritize read transactions.
The selection of transactions to abort is a binary integer programming problem that minimizes the number of aborted transactions.
While these mechanisms enhance performance, they add a little overhead and possibly alter the sequence of transactions within a block compared to the original HLF transaction flow, adding additional complexity and potentially yielding different results.

\section{Proposed Solution} \label{sec:solution}
Our research proposes two distinct modifications of the typical EOV transaction flow model, denoted as \textit{EOV-Original (EOV-OG)}, named OEMVCC and OEMVCC-EA.
These modifications rely on reallocating the MVCC process from the peers to the ordering service.
This adjustment enhances concurrency between the ordering and validation phases, detecting invalid transactions as soon as possible while preserving the same order of transactions that an EOV-OG network would have.

In this section, we will delve deeper into the intricacies of OEMVCC and OEMVCC-EA, discussing their differences and advantages in detail.
We outline their pseudo-code in Algorithms \ref{alg:oemvcc-ordering}-\ref{alg:oemvcc-peer-endorsement}, where an input boolean parameter, named $ea$, is set to differentiate the two mechanisms.
As OEMVCC-EA is an extension of OEMVCC, we will begin by analyzing the logic behind the less intricate inner workings of OEMVCC before examining the more sophisticated OEMVCC-EA.

\subsection{OEMVCC}
\begin{figure}[htbp]
\vspace{-0.2cm}
\centerline{\includegraphics[width=0.5\textwidth]{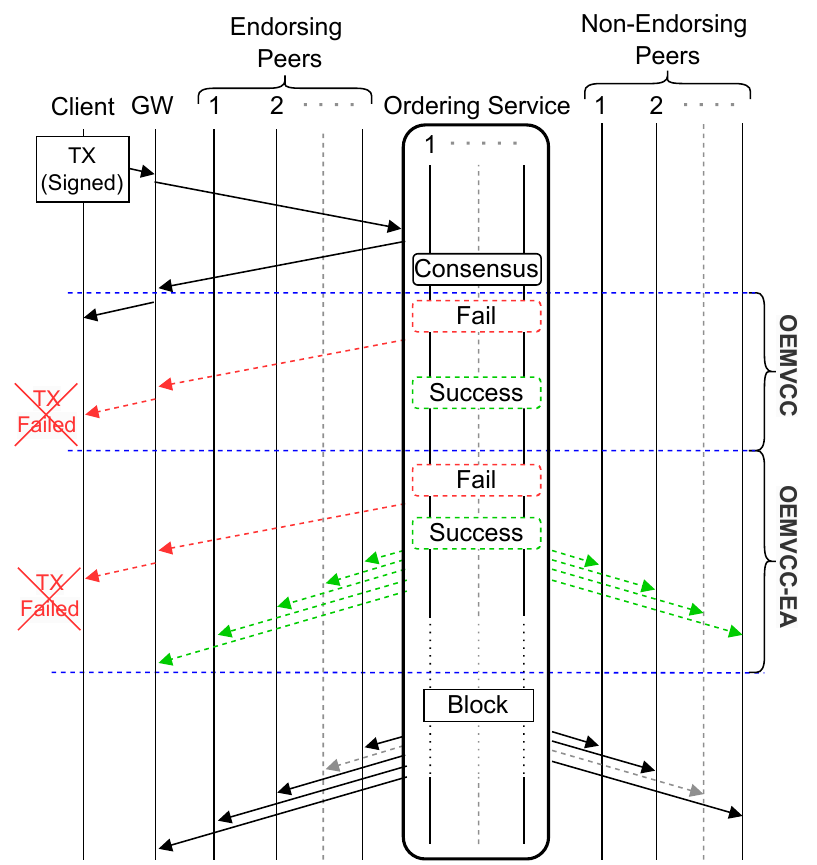}}
\vspace{-0.2cm}
\caption{Transaction Ordering Phase with OEMVCC and OEMVCC-EA.}
\label{fig:ordering}
\end{figure}

Implementing OEMVCC requires modifications across the various components of the HLF network, including the ordering service, the GW, and the peers. 
The most significant modification is the transfer of the MVCC process from the peers to the orderers.
As illustrated in Figure~\ref{fig:ordering} (in OEMVCC) and presented in Algorithm~\ref{alg:oemvcc-ordering} ($ea$ is false), when the ordering service receives a transaction from a GW, it orders it into a block (lines 3-4) using a consensus mechanism.
The orderers then asynchronously perform MVCC (lines 5-6), to classify and mark invalid transactions within a block.
When a transaction is invalid, the ordering service promptly notifies the client that the transaction failed via the GW that sent the transaction (line 11). 
When a block reaches its maximum block size or block interval elapses, the ordering service waits until all transactions within the block have undergone MVCC  before broadcasting it to the peers (lines 20-23).

\begin{algorithm}[t]
\caption{Orderer in OEMVCC or OEMVCC-EA}
\label{alg:oemvcc-ordering}
\begin{algorithmic}[1]
\Function{OEMVCC}{$ea$}
  \While{$true$}
    \State $\text{$tx, gw$} \gets \text{receive()}$
    \State $\text{$block\text{.append(}tx\text{)}$}$
    \State $\text{Start MVCC thread running \{}$
    \State $\ \ \ \ tx.mvcc \gets MVCC(tx)$
    \State $\text{\ \ \ \ } \text{\textbf{if} } tx.mvcc = 0 \text{\textbf{ then}}$
    \State $\text{\ \ \ \ \ \ \ \ } \text{\textbf{if} } ea \text{\textbf{ then}}$
    \State $\text{\ \ \ \ \ \ \ \ \ \ \ \ $block$.remove($tx$)}$
    \State $\text{\ \ \ \ \ \ \ \ \text{\textbf{end if}}}$
    \State $\text{\ \ \ \ \ \ \ \ \text{notify($gw$, $tx$)}}$
    \State $\text{\ \ \ \ \text{\textbf{else if} $ea$ \textbf{then}}}$
      \State $\text{\ \ \ \ \ \ \ \ \text{\textbf{for} $key$ in $tx.wset$} \textbf{do}}$
      \State $\text{\ \ \ \ \ \ \ \ \ \ \ \ \text{\textbf{for} $peer$ in $peers$} \textbf{do}}$
      \State $\text{\ \ \ \ \ \ \ \ \ \ \ \ \ \ \ \ \text{$peer \rightarrow cache[key] \gets 1$}}$
      \State $\text{\ \ \ \ \ \ \ \ \ \ \ \ \text{\textbf{end for}}}$
      \State $\text{\ \ \ \ \ \ \ \ \text{\textbf{end for}}}$
    \State $\text{\ \ \ \ \text{\textbf{end if}}}$
    \State $\text{\}}$
    
    \If{\text{$block\_size \text{ or } block\_interval$}}
      \State $\text{Wait pending MVCC threads and}$  
      \State $\text{broadcast the } block \text{ to peers}$
    \EndIf   
  \EndWhile

\EndFunction
\end{algorithmic}

\end{algorithm}

\begin{figure}[htbp]
\vspace{-0.4cm}
\centerline{\includegraphics[width=0.5\textwidth]{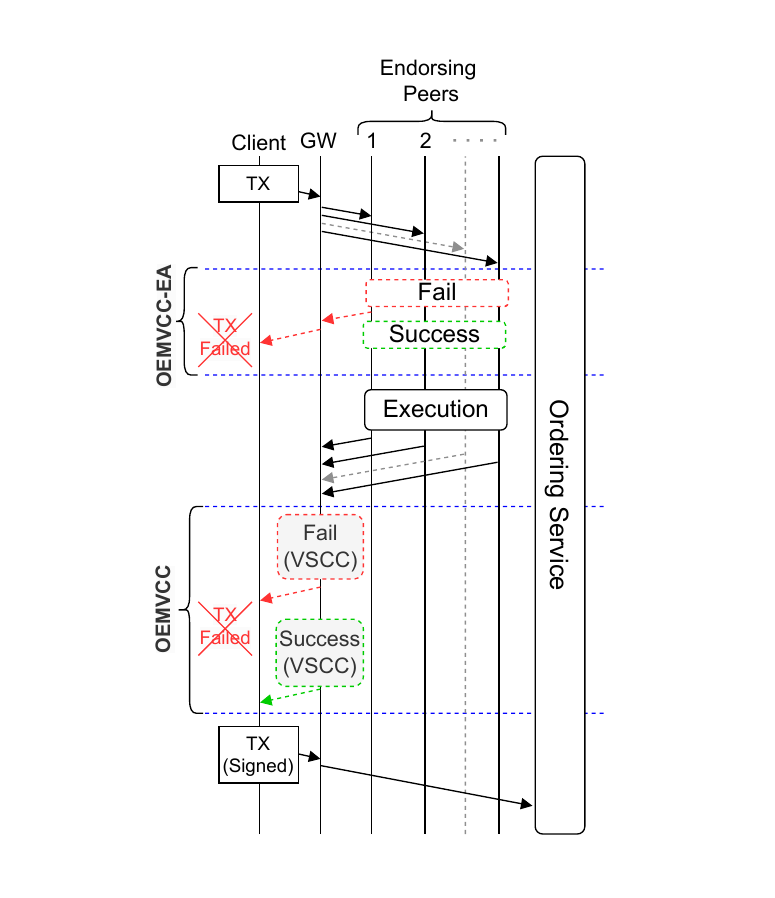}}
\vspace{-0.2cm}
\caption{Transaction Execution Phase.}
\label{fig:execution}
\vspace{-0.2cm}
\end{figure}

OEMVCC takes advantage of the latest versions of HLF, which can incorporate VSCC on the GW, invalidating transactions before dispatching them to the ordering service and notifying the client earlier in the transaction flow, as presented in Figure~\ref{fig:execution} (in OEMVCC).
Performing VSCC directly after the execution phase on the GW increases the probability that valid MVCC transactions on orderers will successfully pass VSCC on the validation phase and eventually change the version of their keys upon block commitment.
Peers must still perform VSCC on the validation phase to avoid malicious clients that might forge the results of the execution phase and submit the transaction directly to the ordering service. 
   
OEMVCC implements a cache-based mechanism to incorporate MVCC and avoid direct access to the world state on the ordering service, described in Algorithm \ref{alg:oemvcc-mvcc}.
The MVCC procedure, called in Algorithm \ref{alg:oemvcc-ordering} (line 6), checks the version numbers of the read-sets for each transaction against the cache (lines 4-10).
A transaction is invalid when the version of at most one of its keys is outdated compared to the one recorded in the cache (lines 5-8).
All keys with the correct version get locally stored until all keys involved in the transaction are verified against the cache (line 9).
If all stored keys succeed in the MVCC check, the transaction maintains its validity, updating the cache with their expected versions upon commitment (lines 11 - 17).

\begin{algorithm}
\caption{MVCC in the Ordering Service}
\label{alg:oemvcc-mvcc}

\begin{algorithmic}[1]

\Function{MVCC}{$tx$}
    \State $tx.mvcc \gets 1$
    \State $updates \gets [\ ]$
    \For{$key$, $version$ in $transaction.rset$}
        \If{\text{$key \in cache \text{ and } version < cache[key]$} }          
          \State $tx.mvcc \gets 0$
          \State break
        \EndIf   
        \State $updates.\text{append(}\{key:version\}\text{)}$
    \EndFor
    
    \If{\text{$tx.mvcc = 1$}}
      \For{$key$, $version$ in $updates$}
        \For{$orderer$ in $orderers$}
          \State $orderer \rightarrow cache[key] \gets version + 1$
        \EndFor
      \EndFor
    \EndIf       
\EndFunction
\end{algorithmic}
\end{algorithm}

The caching mechanism within the ordering service should account for the logic of the underlying consensus algorithm.
Raft regularly elects a leader orderer who orchestrates the ordering of transactions into blocks for its leadership duration.
Followers forward incoming transactions to their leader, ensuring strong consistency between their log entries and those of the leader.
In OEMVCC, the leader also performs MVCC, indicating that the caching mechanism should incorporate strong consistency in its entries to accommodate elections, as a follower can transition into a leader at any time and undertake the ordering and MVCC.
Each orderer maintains one write-through cache for each channel it belongs to, ensuring privacy and isolation between channels.
The leader updates the caches of its followers after it orders a transaction into a block and before ordering subsequent transactions.
When the leader crashes or when a new leader is elected before the caches are updated, the new leader is responsible for updating the caches according to its log entries or the incomplete block received.
Finally, if the caches become full, the orderers transfer the oldest entries to persistent storage.

\begin{figure}[htbp]
\vspace{-0.4cm}
\centerline{\includegraphics[width=0.5\textwidth]{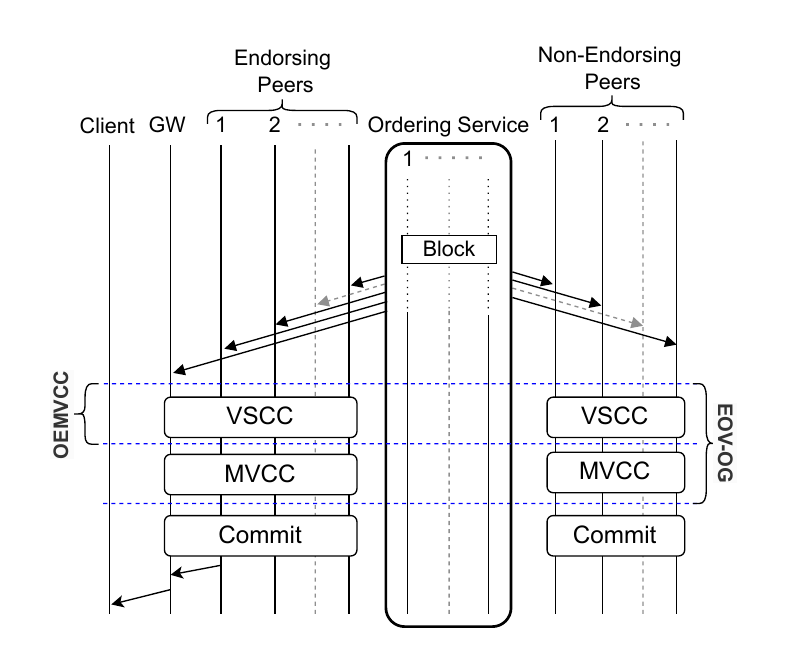}}
\caption{Transaction Validation Phase.}
\vspace{-0.2cm}
\label{fig:validation}
\end{figure}

The behavior of peers during the validation phase also depends on whether OEMVCC or OEMVCC-EA is employed, as outlined in Algorithm \ref{alg:oemvcc-peer-validation}.
Contrary to the EOV-OG, where transactions sequentially undergo VSCC and MVCC, upon receiving a block, peers in OEMVCC, depicted in Figure~\ref{fig:validation} (in OEMVCC), solely perform VSCC to detect transactions submitted by malicious clients, which bypass VSCC on the GW (lines 5-9). 
After VSCC, the peer can commit the block directly to the ledger (line 16), avoiding MVCC. 
Finally, we extend the GW to support receiving and handling notifications by the ordering service to notify the client regarding invalid transactions.

\begin{algorithm}
\caption{Peer Validation in OEMVCC or OEMVCC-EA}
\label{alg:oemvcc-peer-validation}

\begin{algorithmic}[1]
\Function{VALIDATE}{$ea$}
  \While{$true$}
    \State $block \gets \text{receive()}$
    \For{$tx$ in $block$}
    \If{$not\ ea$}
        \State $tx.vscc \gets VSCC(tx)$
        
          \If {$tx.vscc = 0\text{ OR }tx.mvcc = 0$}
            \State continue
          \EndIf       
        
      \Else
        \For{$key$ in $tx.wset$}
            \State $peer \rightarrow cache[key] \gets 0$
        \EndFor
    \EndIf  
    \EndFor

    \State $\text{Commit the block to the ledger}$  
  \EndWhile
\EndFunction
\end{algorithmic}
\end{algorithm}

\subsection{OEMVCC-EA}
OEMVCC-EA extends upon OEMVCC, aiming to increase transaction throughput while conserving network and computational resources by dropping invalid transactions earlier in the transaction flow.
Figure~\ref{fig:ordering} (in OEMVCC-EA) and Algorithm \ref{alg:oemvcc-ordering} ($ea$ is true) describe how the ordering service incorporates OEMVCC-EA.
OEMVCC-EA assumes no malicious clients exist on the network and performs VSCC exclusively on the GW.
Since the GW and the ordering service detect invalid VSCC and MVCC transactions, block sizes can be reduced by discarding invalid transactions after the ordering phase (lines 8-10), creating blocks that contain solely valid transactions, and notifying the clients earlier through the GW (line 11).
The ordering service proactively informs the peers about the keys in the write sets of valid transactions to optimize their execution and validation phases (lines 12-18).

\begin{algorithm}[htbp]
\caption{Peer Execution in OEMVCC or OEMVCC-EA}
\label{alg:oemvcc-peer-endorsement}

\begin{algorithmic}[1]
\Function{ENDORSE}{$ea$}
  \While{$true$}
    \State $\text{$tx, gw$} \gets \text{receive()}$
    \State $\text{$tx.mvcc$} \gets 1$

    \If{$ea$}

      \For {$key$ in $tx$}
        \If{$peer \rightarrow cache[tx.key] = 1$}
          \State $\text{$tx.mvcc$} \gets 0$
          \State $\text{break}$
        \EndIf
      \EndFor
    \EndIf
      
    \If{$tx.mvcc = 0$}
          \State $\text{notify($gw$, $tx$)}$
    \Else
      \State $\text{Simulate the execution of the transaction}$  
        
    \EndIf
  \EndWhile
\EndFunction
\end{algorithmic}
\end{algorithm}

Endorsing peers utilize the write sets of valid transactions to avoid simulating transactions containing these keys since the resulting read-write sets will eventually become outdated, and the transaction will ultimately fail, as described in Figure~\ref{fig:execution} (in OEMVCC-EA) and Algorithm \ref{alg:oemvcc-peer-endorsement} ($ea$ is true) (lines 3 - 14).
Finally, as the peers receive blocks containing exclusively valid transactions, they can directly commit the included transactions, as illustrated in Figure~\ref{fig:validation} (in OEMVCC-EA) and described in Algorithm \ref{alg:oemvcc-peer-validation}.
Once a transaction is committed to the ledger by a peer, the peer can remove the keys included on the write of the validated transaction (lines 11-13), enabling it to endorse upcoming transactions that involve these keys.

\section{Evaluation Framework} \label{sec:evaluation}
To evaluate the performance of our proposed modifications, including OEMVCC and OEMVCC-EA, we deployed an HLF network based on HLF v2.4.4 on the NITOS \cite{nitos} testbed, extending upon the predefined deployment options detailed in \cite{fabric-samples}. 
In particular, the network consists of ten clients running their separate client applications, three orderers within the ordering service, and four peer nodes.
The ordering service operates from a single NITOS node, with all orderers deployed on the same NITOS node, incorporating the default Raft consensus mechanism for ordering and a block size of 10 with a block interval of 2 seconds.
We also deploy the clients alongside the peers, hosted on distinct NITOS nodes, separately from the ordering service.
Finally, for the distributed caching mechanism used in peers and orderers, we use a distributed Redis cache \cite{redis}.

\subsection{Experiments Overview}
Anticipating that our modifications will predominantly enhance the performance in scenarios containing high rates of conflicting transactions, we enhanced the existing \textit{asset transfer} \cite{fabric-samples} client application.
Our focus is for clients to concurrently submit conflicting transactions that attempt to modify similar assets to investigate the improvements in performance relative to the conflict rate.
Clients randomly select an asset to modify from an asset pool based on a rate of conflicting transactions.
Each then randomly selects two out of the four available peers to send its transaction proposals for execution, adhering to an endorsement policy that requires at least one endorsement before proceeding to the ordering phase.

In our experiments, we assess the efficiency of EOV-OG alongside OEMVCC and OEMVCC-EA, focusing on the average transaction throughput and latency.
For each, the clients concurrently submit varying levels of conflicting transactions, including $20\%$, $50\%$, and $80\%$ of conflict rates. 
The experimental results determine how they perform under various transaction conflict rates and quantify their impact on the overall performance and variability.

\subsection{Experimental Results}
Figure~\ref{fig:endorsement-duration} illustrates the relationship between average transaction execution duration during the execution phase and the percentage rate of conflicting transactions.
Figure~\ref{fig:latency} depicts the average overall transaction latency, measured from when a transaction gets submitted by a client until the client receives its status after the transaction has been committed to the ledger.
Figure~\ref{fig:throughput} showcases the average overall transaction throughput, measured by the volume of notifications sent to clients regarding valid and invalid transactions within a given time frame.
To provide a clear comparison of the performance of the different modifications, we denote the OEMVCC with a blue horizontally-lined bar, the OEMVCC-EA with a red vertically-lined bar, and the EOV-OG with a black diagonally-lined bar.
Furthermore, the average latency and throughput of transactions, including the overall, valid, and invalid transactions, are represented by the circle-pointed, square-pointed, and triangle-pointed symbols, respectively.
By analyzing these measurements, we can assess the advantages of our proposed modifications alongside their shortcomings, as they indirectly provide insights into the load on the peers and demonstrate the overhead associated with our solution.
To delve deeper, we will first compare the performance of OEMVCC and OEMVCC-EA with the EOV-OG.
Subsequently, we will conduct a comparative analysis between OEMVCC and OEMVCC-EA.

\subsubsection{OEMVCC and OEMVCC-EA vs EOV-OG}

\begin{figure}[htbp]
\centerline{\includegraphics[width=0.5\textwidth]{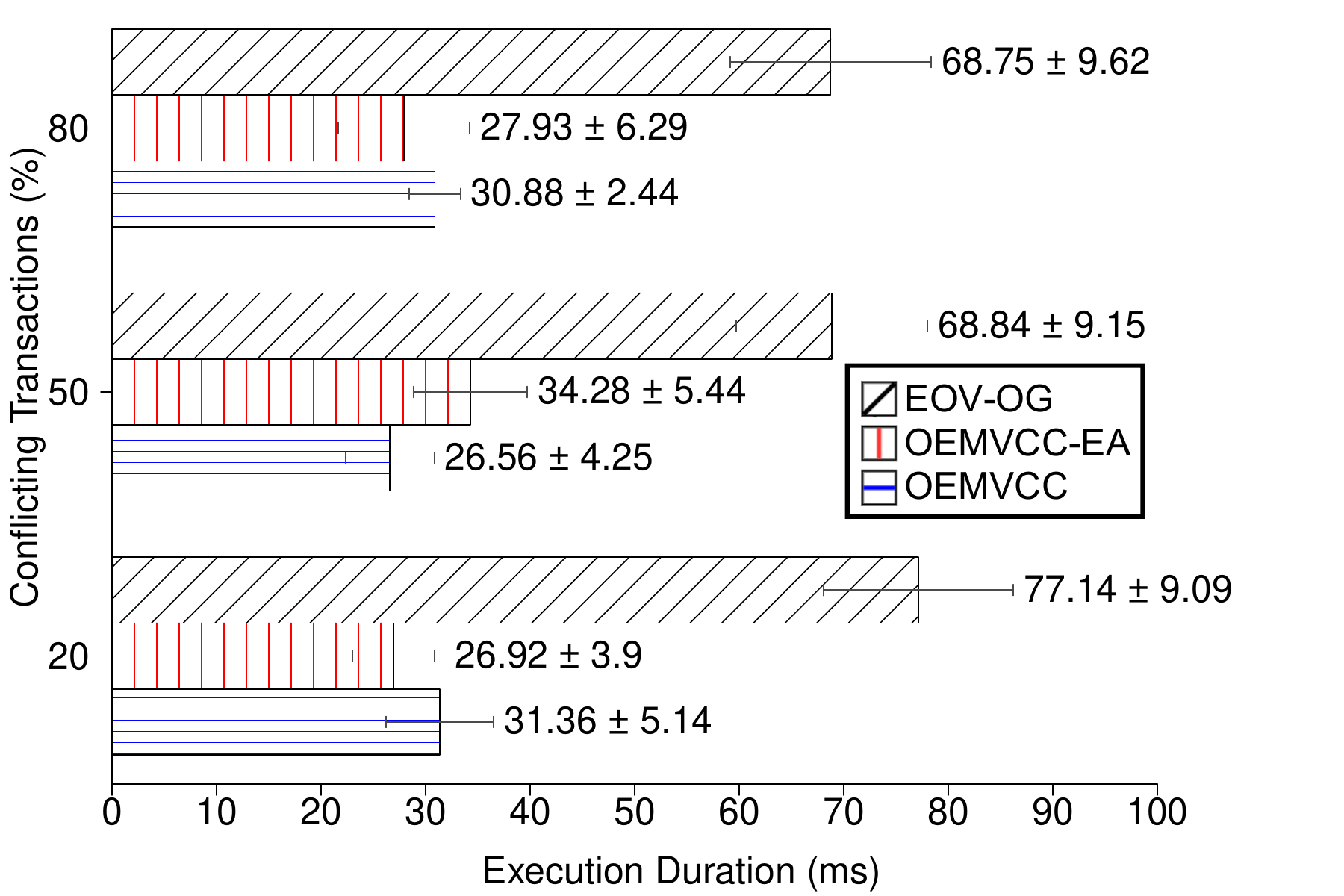}}
\vspace{-0.2cm}
\caption{Execution duration relative to the rate of conflicting transactions.}
\label{fig:endorsement-duration}
\end{figure}

In Figure~\ref{fig:endorsement-duration}, we observe that OEMVCC and OEMVCC-EA significantly outperform EOV-OG, slashing the execution duration by at least half, irrespective of the conflict rate. 
Given that the execution duration closely mirrors the load on the peers, this substantial reduction in execution time for OEMVCC and OEMVCC-EA underscores the efficient management of MVCC within the ordering service and the significance of invalidating invalid transactions as soon as possible.
By offloading MVCC to the ordering service and invalidating invalid transactions earlier in the transaction flow, the peers can conserve resources, enabling more efficient transaction simulation during the execution phase.

\begin{figure}[htbp]
\centerline{\includegraphics[width=0.5\textwidth]{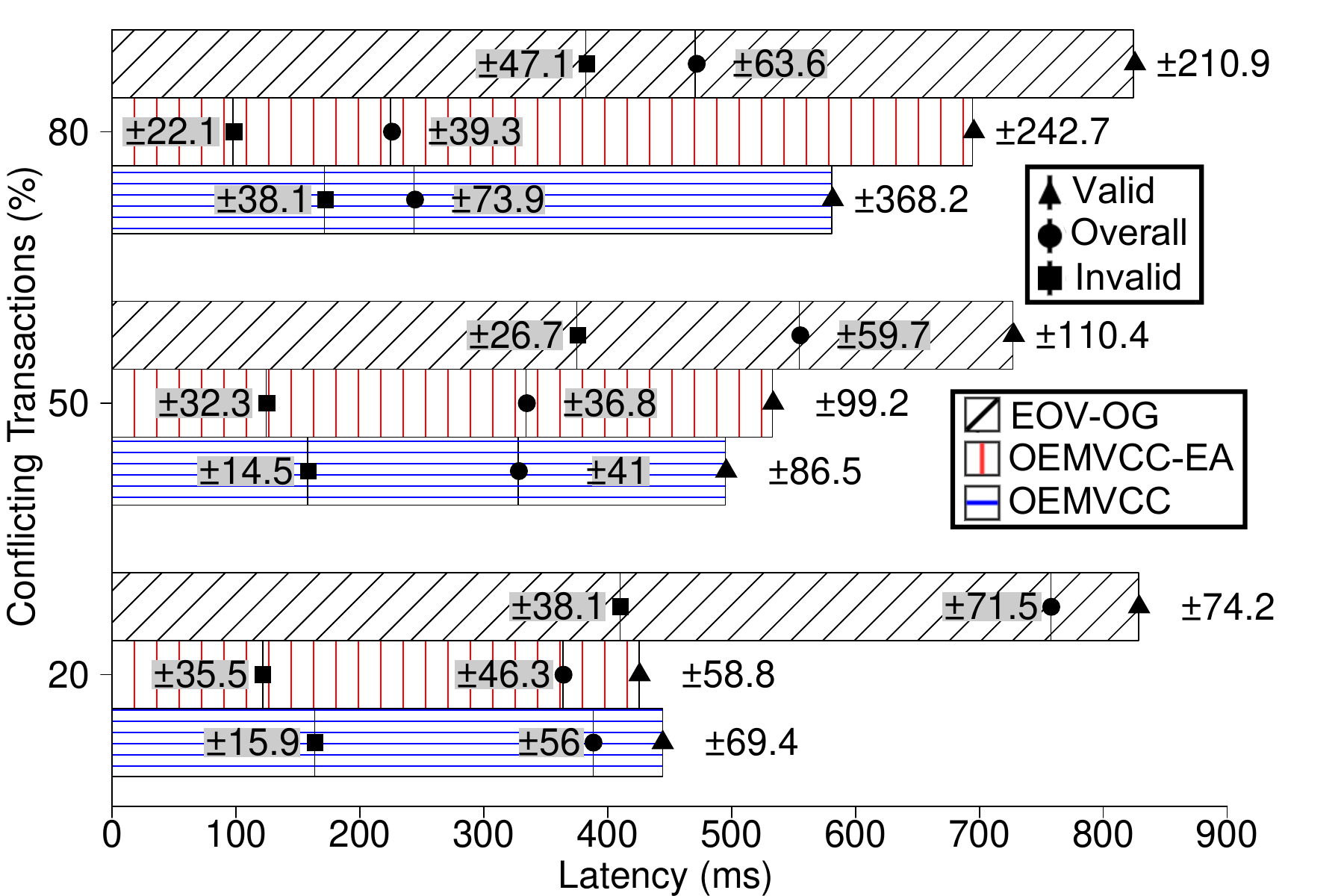}}
\vspace{-0.2cm}
\caption{Transaction latency relative to the rate of conflicting transactions.}
\label{fig:latency}
\end{figure}

Similarly, the transaction latency of OEMVCC and OEMVCC-EA is substantially lower than that of EOV-OG, outperforming it regardless of the transaction conflict rate, as illustrated in Figure~\ref{fig:latency}.
This reduction in the overall latency stems from the simultaneous latency reduction of invalid and valid transactions.
Invalid transactions are identified at earlier stages in the transaction flow, allowing prompt notification about each invalid transaction to the client that submitted it and subsequently lower latency. 
Valid transactions, conversely, leverage the benefits of asynchronous MVCC and transaction ordering in the ordering service. 
This approach bypasses the sequential execution of ordering and validation phases found in EOV-OG, reducing the latency of valid transactions.
Moreover, as outlined earlier in the explanation of the execution duration reduction, the transfer of MVCC from peers to the orderers and the earlier invalidation of transactions, particularly in OEMVCC-EA, enables peers to conserve more resources for processing valid transactions. 
This conservation of resources contributes significantly to the reduced latency of valid observed.

\begin{figure}[htbp]
\centerline{\includegraphics[width=0.5\textwidth]{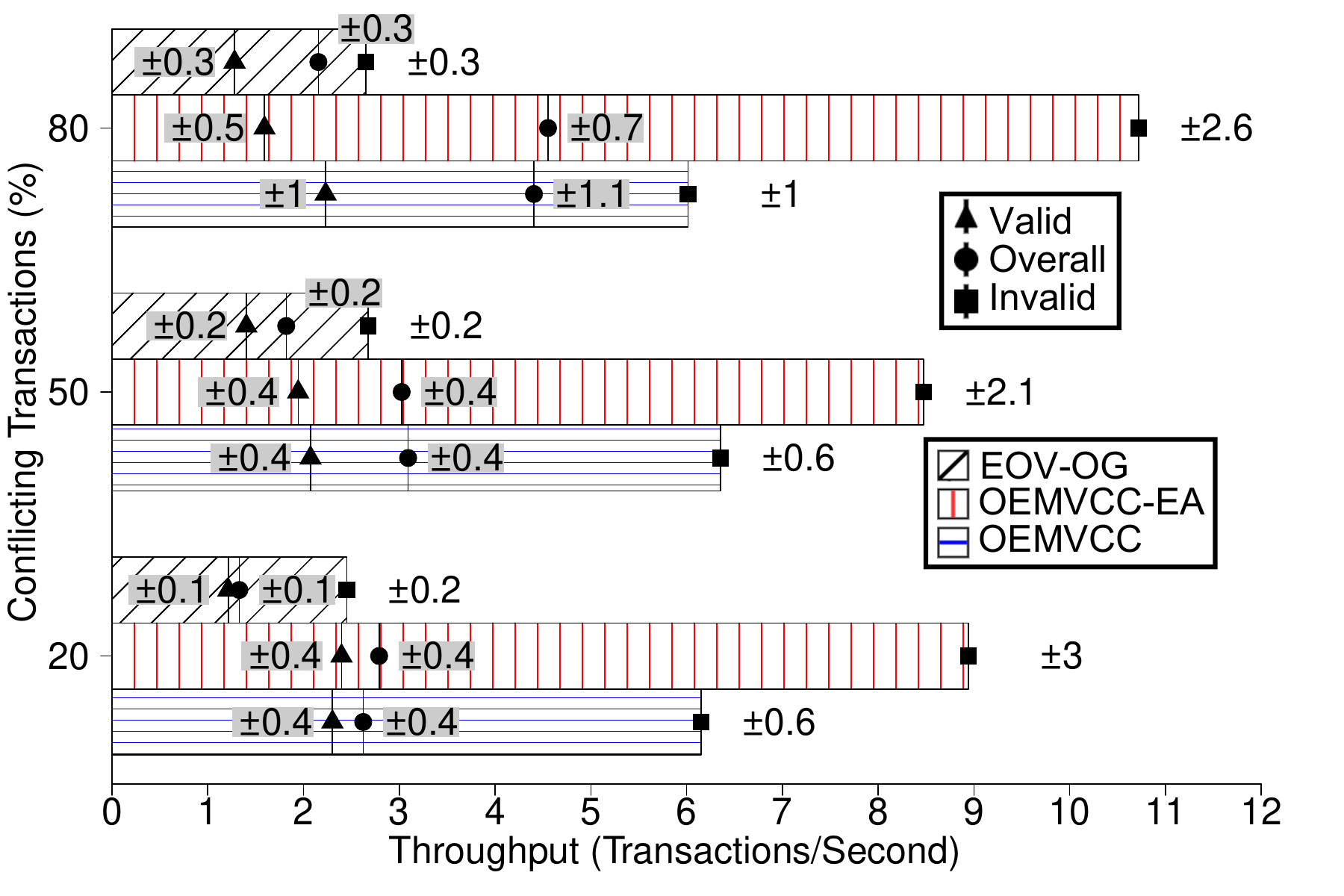}}
\vspace{-0.2cm}
\caption{Transaction throughput relative to the rate of conflicting transactions.}
\label{fig:throughput}
\end{figure}

Finally, the transaction throughput of OEMVCC and OEMVCC-EA surpasses that of EOV-OG, consistently outperforming it across various transaction conflict rates, as demonstrated in Figure~\ref{fig:throughput}.
This superiority in throughput arises from the lower latency observed in transactions, including overall, valid, and invalid transactions, in OEMVCC and OEMVCC-EA compared to EOV-OG. 
Consequently, the average transaction throughput is higher for OEMVCC and OEMVCC-EA than for EOV-OG.
Additionally, the standard deviation of OEMVCC and OEMVCC-EA is higher than that of EOV-OG as they heavily depend on the probability of early invalidation of invalid transactions. 

\subsubsection{OEMVCC vs OEMVCC-EA}
The execution duration, average overall latency, and throughput of OEMVCC-EA excel over OEMVCC at $20\%$ and $80\%$ conflict rates, which is not the case for $50\%$.
The advantage of OEMVCC-EA arises as invalid transactions get invalidated earlier in the transaction flow, in the execution phase, while also conserving resources for possibly valid transactions.
However, the overhead of the cache-based nature of OEMVCC-EA emerges at equal ratios of non-conflicting and conflicting transactions, denoted as the $50\%$ conflict rate.
In OEMVCC-EA, peers not only commit transactions but simultaneously update the status of the keys in those transactions within their cache, used in the execution phase to avoid unnecessary simulation of transactions that would eventually fail.
This simultaneous activity can impede the peers and the overall performance.
At $20\%$ conflict rates, the performance improvement achieved by resource conservation outweighs the intense resource utilization associated with the frequent application of OEMVCC-EA on the peers.
For conflict rates as high as 80\%, this concurrent activity similarly impacts the execution phase, albeit its effects show to a lesser extent, as more transactions get proactively invalidated at such high conflict rates, conserving sufficiently enough resources within the network to improve the overall performance.
However, at conflict rates of around 50\%, the overhead of OEMVCC-EA offsets the benefits of resource conservation, resulting in a decline in performance.
This equilibrium of overhead and resource conservation is the reason why the endorsement duration in Figure~\ref{fig:endorsement-duration}, the overall latency in Figure~\ref{fig:latency}, and the overall throughput in Figure~\ref{fig:throughput} are all underperforming in OEMVCC-EA at a $50\%$ conflict rate compared to OEMVCC.

The advantage of OEMVCC-EA becomes apparent in handling invalid transactions.
As Figure~\ref{fig:latency} depicts, OEMVCC-EA outperforms OEMVCC in the latency of solely invalid transactions as it can detect and invalidate invalid transactions sooner in the transaction flow than OEMVCC.
However, the average latency of valid transactions in OEMVCC-EA exceeds that of OEMVCC under $50\%$ and $80\%$ conflict rates, while it is lower at a $20\%$ conflict rate due to the overhead of OEMVCC-EA, especially showcased in valid transactions.
Similarly, in Figure~\ref{fig:throughput}, we observe that the throughput of invalid transactions in OEMVCC-EA is greater than that of OEMVCC, regardless of the conflict rate.
Nevertheless, the overhead of OEMVCC-EA increases with the percentage of conflict rate, reducing the average throughput of valid transactions.
Due to this overhead, valid transactions perform better in OEMVCC compared to OEMVCC-EA under $50\%$and $80\%$ conflict rates.
In contrast, valid transactions perform worse in OEMVCC than OEMVCC-EA in lower conflict rates of $20\%$, as the overhead is minimal.
Finally, OEMVCC-EA exhibits a higher standard deviation in the throughput of invalid transactions, primarily due to its heavy reliance on the timeliness of information within the caches of peers, which lack strong consistency.

\section{Conclusions \& Future Work} \label{sec:conclusion}
Our research demonstrates that OEMVCC and OEMVCC-EA exhibit substantial performance advantages over EOV-OG in transaction latency and throughput, regardless of the conflict rate of the client applications. 
Specifically, OEMVCC-EA shows distinct advantages in handling invalid transactions, owing to its ability to detect and invalidate invalid transactions earlier than OEMVCC, achieving superior performance in high-conflict applications.
Additionally, OEMVCC-EA performs better in low-conflict applications where its overhead is minimal.
Conversely, OEMVCC excels in managing valid transactions when applications submit equal ratios of non-conflicting and conflicting transactions, where the overhead of OEMVCC-EA becomes significant.

For future work, we will integrate and evaluate OEMVCC and OEMVCC-EA in realistic applications that feature variable conflict rates according to their requirements. 
Additionally, we aim to introduce a mechanism to dynamically switch between these two modifications based on the rate of conflicting transactions to achieve the best possible performance.
Finally, we intend to enhance the efficiency and scalability of OEMVCC and OEMVCC-EA in HLF deployments with high latency between nodes.

\section*{Acknowledgments}
This project has received funding from the European Union’s HE research and innovation programme under grant agreement No 101084188. Views and opinions expressed are, however, those of the authors only and do not necessarily reflect those of the European Union or the European Research Executive Agency (REA). Neither the European Union nor the granting authority can be held responsible for any use that may be made of the information the document contains.

\bibliographystyle{unsrt}
\bibliography{IEEEabrv,bibfile}

\end{document}